\documentclass[10pt,twocolumn,twoside]{IEEEtran}

\usepackage{amssymb,amsfonts}
\usepackage{eucal,bm}
\usepackage{gensymb,subfigure}
\usepackage{booktabs}
\usepackage{times,color,url}
\usepackage{comment}

\usepackage{cite}
\usepackage{graphicx}
\usepackage[cmex10]{amsmath}

\begin{document}
\title{Optimal compression in natural gas networks: a geometric programming approach}

\author{Sidhant~Misra, Michael~W.~Fisher, Scott~Backhaus, Russell~Bent, Michael~Chertkov, Feng Pan
\thanks{S. Misra, M.W. Fisher and M. Chertkov are with Theory Division of LANL, Los Alamos, NM 87544}
\thanks{S. Misra is also with Department of Electrical Engineering and Computer Science, MIT, Cambridge, MA, 02139}
\thanks{M. W. Fisher is also with Department of Electrical Engineering and
Computer Science, University of Michigan, Ann Arbor, 48105}
\thanks{S. Backhaus is with MPA Division of LANL, Los Alamos, NM 87544}
\thanks{R. Bent and F. Pan are with DSA Division of LANL, Los Alamos, NM 87544}

\thanks{Manuscript received September 20, 2013}}

\markboth{IEEE Transactions on Control of Network Systems (CONES)}
{Misra \MakeLowercase{\textit{et al.}}: Optimal compression in natural gas networks}

\maketitle

\begin{abstract}
Natural gas transmission pipelines are complex systems whose flow characteristics are governed by challenging non-linear physical behavior. These pipelines extend over hundreds and even thousands of miles.  Gas is typically injected into the system at a constant rate, and a series of compressors are distributed along the pipeline to boost the gas pressure to maintain system pressure and throughput.  These compressors consume a portion of the gas, and one goal of the operator is to control the compressor operation to minimize this consumption while satisfying pressure constraints at the gas load points. The optimization of these operations is computationally challenging. Many pipelines simply rely on  the intuition and prior experience of operators to make these decisions.  Here, we present a new geometric programming approach for optimizing compressor operation in natural gas pipelines.  Using models of real natural gas pipelines, we show that the geometric programming algorithm consistently outperforms approaches that mimic existing state of practice.
\end{abstract}

\begin{IEEEkeywords}
Natural Gas Network, Optimal Compression,  Geometric Programming, Dynamic Programming
\end{IEEEkeywords}


\section{Introduction: History \& Motivation}

\IEEEPARstart{I}{n} recent years, worldwide natural gas reserves have expanded at a rapid pace.  The invention and application of hydraulic fracturing in the US has enabled the economic capture of many sources of unconventional natural gas \cite{10CWB} while improved exploration techniques and increased offshore activity has led to increased conventional reserves in several countries.  The increased availability and lower cost of gas in these regions are making it more attractive economically.  In the US, the economic advantage of gas is pushing out coal (and to a lesser extent fuel oil) as a primary source of energy.  In addition, the lower CO2 emissions from gas mitigate much of the uncertainty related to the future economic cost of carbon emissions. These properties make gas a very attractive bridge fuel to a low carbon economy, and this shift is already occurring in several regions of the US electric sector \cite{2010MITEI}.  The high cost and long economic lifetime of the electrical generation assets acts to lock in this shift to a large degree.

The cost of the fuel is not the only advantage of natural gas over coal and fuel oil.  From the planning and construction point of view, the physical footprint and total emissions of gas turbines is smaller than coal or fuel oil-based generation easing the difficulty of siting and permitting.  From an operational perspective, gas turbines can quickly change their generation output in response to changes in intermittent renewable generation such as wind.  This ability to move quickly is also manifest in the ability to quickly start up a gas turbine from a cold condition (especially single-cycle gas turbines).  The combination of these benefits is driving the higher penetration of gas turbines into the electrical grid.  The Independent System Operator of New England (ISO-NE) is a prime example.  Over the two decades, the level of gas generation in ISO-NE has increased from 5\% to 50\% of total generation capacity \cite{ISO-NE}.
However, the benefits of natural gas are not without some risk.  As the level of natural gas-based generation increases, larger and perhaps more variable,\footnote{Natural gas generation is often used to smooth the variability in renewable energy sources.} natural gas loads will effect the operations of the large natural gas transmission pipelines that bring the gas from the sources to the generator and other gas loads. The impact is not just one-way.  The finite capacity of these gas transmission pipelines will limit the availability of gas which will directly affect ability of natural gas generators to respond to grid operator control commands.

The majority of the distance between gas sources and gas generators and other loads is covered by large, high-pressure transmission pipelines.  High pressure and density enable high throughput with the pressure drop driving the gas through the pipeline.  As the pressure falls, the flow velocity increases (under constant mass flux) and the pressure then falls even faster.  Gas compressors are used to maintain the throughput of pipeline and maintain the required pressure at the customer load points.  Often these gas compressors are driven by gas engines that burn natural gas from the pipeline itself.  Typical designs of transmission pipelines places compressors every $\sim$50-100 miles.  In large transmission pipelines that span 600 miles or more, compressors  consume (burn)  $\sim2-5\%$ of the transmitted gas.  This burn is a cost of transporting the gas, and who bears that cost affects the goals of the operational optimizations (discussed below). Complicating the domain, the bearer of this cost differs from country to country.

The difficulty and expense of building new or expanding large-scale infrastructure coupled with the increasing (and the potentially more time-variable) gas loads calls for improved optimization of pipeline operations.  However, the goals of these optimizations must be aware of and developed within the regulatory, market and ownership frameworks of the pipelines. Here, we briefly review two existing frameworks that are at opposite ends of this regulatory/ownership spectrum.  Norway presents a relatively simple framework.  In Norway, gas sources, gas pipelines, and the sale of gas inside and outside the country is controlled by the government.  Norway produces more gas than can be domestically consumed and has strong economic motivations to sell this excess to the rest of Europe.  The demand for the gas (and the available Norwegian gas resource) is typically higher than the ability of Norway's pipeline network to transport the gas to markets at its border.  To increase sales and revenues, the pipeline operator's primary objective is to increase the pipeline {\it throughput}, and the gas lost to compression offsets improvements in throughput making the optimal compression problem important in this context.  The throughput on the Norwegian is complicated by the differing gas compositions required by the buyers of the gas and the differing compositions of the gas sources.  See \cite{10Bor} for a discussion of this problem.

In the US, gas markets have been deregulated for many years \cite{2010MITEI}.
The implication is that pipeline operators do not own sources of gas nor are they involved in sourcing and selling gas to consumers (gas distribution companies, industrial consumers, or gas turbine generators).  Instead, the pipeline operators are responsible for transporting the gas and maintaining and expanding the pipelines.  Gas is sold in organized markets via bi-lateral arrangements between gas suppliers and consumers.  In addition to securing the gas itself, the consumers (buyers) must have also purchased the right to move the gas though the pipeline from the gas sources to the gas load locations.  It is the sale of these rights where pipeline owner/operators make their revenue, and reliably increasing the throughput of the pipeline can enable the owner/operator to secure additional revenue.  Therefore, as with the case of Norway, the US pipeline operators have an interest in increasing the pipeline {\it throughput}.  Gas lost to compression offsets improvements in throughput making the optimal compression problem important in this context.

Within these disparate pipeline ownership/operational frameworks, minimizing the cost of compression is an important problem whose solution will enable additional pipeline throughput.  Throughput could also be improved through the optimal placement of new compressors, however, here we focus on the optimal operation of existing compressors.  The early compression cost minimization model was solved by Dynamic Programming (DP) and can be traced back to \cite{68WL}.  An excellent review of the literature on compression cost minimization can be found in \cite{10Bor}. The key contribution of this paper is the development of a Geometric Programming (GP) based approach for optimizing the transport of natural gas.  It offers optimality properties similar to existing algorithms reviewed in \cite{10Bor}, however it is a convex optimization approach which offers desirable convergence properties without the need for discretization. We focus on developing GP for steady-state gas flow models on tree networks. Given existing engineering practices and network design, these are natural assumptions. However, it is important to note that GP potentially has several advantages when considering extensions to the problem that are expected to be needed in the future. These features include stochastic gas draws, loops, distributed control, risk mitigation, transient dynamics, and interdependencies with power systems. In these cases, the GP formulation has natural mechanisms for incorporating these features that are unavailable to DP. These extensions will be addressed in future work. In this manuscript, we establish that GP matches the performance of existing algorithms in order to motivate its use in more complex settings where existing algorithms are not easily adapted.

The remainder of this manuscript is organized as follows.  Section~\ref{sec:tech_intro} reviews the pipeline gas flow equations and the Optimal Gas Flow (OGF) problem.  Section~\ref{sec:tech_opt}  describes our GP formulation for tree-like gas pipelines. For comparison, we also formulate a Dynamic Programming (DP) approach to the same problems.  Section~\ref{sec:exp} describes the implementation of the GP and DP algorithms as well as a greedy algorithm that is intended to represent how many US pipelines are operated today.  This section also compares the results of applying these approaches to a model of the Belgian natural gas network and the Transco pipeline network in the US \cite{Transco}.  Finally, Section~\ref{sec:conclusions} provides some conclusions and a discussion of potential future research for both the steady-state gas flow problem and the time variable flow (line-packing) problem.

\section{Technical Introduction}
\label{sec:tech_intro}

In this section, we review the gas flow equations and simplifying approximations used by practitioners.  We start from a model of a single pipe, generalize the equations to a network of pipes, and close by embedding the equations in an optimization problem.

\subsection{Gas Flow Equations: Individual Pipe}

To introduce notation and the fundamental physics of gas systems, we first consider the flow of a compressible gas in a single section of pipe. Transmission pipelines are typically 16-48 inches in diameter and operate at high pressures and mass flows, e.g. $200$ to $1500$ pounds per square inch (psi) and  move millions of cubic feet of gas per day \cite{Ref_Crane1982,Ref_Mokhatab2006}. Under these highly turbulent conditions, the pressure drop and energy loss due to shear is represented by a phenomenological friction factor, and the resulting gas flow model is a partial differential equation (PDE) with one spatial dimension $x$ (along the pipe axis) and one time dimension \cite{osiadacz1987simulation,87TT,05Sar}:

\begin{eqnarray}
&& \partial_t\rho+\partial_x (u\rho)=0,\quad p=\rho Z R T\label{density_eq}\\
&& \partial_t (\rho u)+\partial_x (\rho u^2)+\partial_x p=-\frac{\rho u |u|}{2D} f-\rho g \sin\alpha,\label{momenta_eq}
\label{state_eq}
\end{eqnarray}
Here, $u,p,\rho$ are velocity, pressure, and density at the position, $x$; $Z$ is the gas compressibility factor; $T$ is the temperature; $R$ is the gas constant; $D$ is the diameter of the pipe and
 $\alpha$ is its tilt angle; $f$ is the friction factor and; $g$ is the acceleration due to gravity.

Eqs.~(\ref{density_eq},\ref{momenta_eq}) represent mass conservation, the ideal gas thermodynamic relation and momentum balance, respectively. The first term on the rhs of Eq.~(\ref{momenta_eq}) represents the friction losses created in a pipe of diameter $D$ with friction factor $f$.  The second term on the rhs of Eq.~(\ref{momenta_eq}) accounts for the gain or loss of momentum due to gravity $g$ if the pipe is tilted by angle $\alpha$.  The frictional losses typically dominate the gravitational term, which is typically dropped.  Similarly, the gas inertia term, $\partial_t(\rho u)$, and the advection term, $\partial_x (\rho u^2)$, are  typically small compared to the frictional losses and are dropped. For simplicity of presentation, we have also assumed that the temperature does not change significantly along the pipe. In case of long pipes, where temperature gradients do appear, this problem can be resolved by representing the pipe as a series of shorter pipes, each with negligible temperature gradients along their lengths.

Taking into account these assumptions Eqs.~(\ref{density_eq},\ref{momenta_eq}) are rewritten in terms of the pressure $p$ and the mass flux $\phi=u\rho$:
\begin{eqnarray}
&& \partial_t p=-ZRT\partial_x \phi,\label{density_eq1}\\
&& \partial_x p^2=- \frac{f Z R T}{D} \phi |\phi|.\label{momenta_eq1}
\end{eqnarray}
If the flow into and out of the pipe at the two ends balance such that the total mass of gas in the pipe does not change, the flow is steady and Eqs.~(\ref{density_eq1},\ref{momenta_eq1}) can be solved (by setting the time derivatives to zero):
\begin{eqnarray}
\phi=\mbox{const},\quad p_{in}^2-p(x)^2=a \frac{x}{L} \phi |\phi|,\quad a\equiv\frac{f Z R T L}{D}.\label{steady}
\end{eqnarray}
Here , $0\leq x\leq L$, and $L$ is the length of the pipe.  The constant $a$ characterizes the pressure drop due to flow in the pipe and is the only important pipe parameter in the steady-state model.

\subsection{Steady Gas Flow over Network}
The solution in Eq.~(\ref{steady}) is now used to derive a node-edge network model for the case of steady flow.
To continue the discussion, we first consider a Gas Flow (GF) network without compressors which is represented by a directed graph ${\cal G}=({\cal V},{\cal E})$ with edges ${\cal E}$ and vertexes ${\cal V}$. A solution of the steady gas flow problem consists of finding a set of node pressures $p=(p_i\geq 0|i\in{\cal V})$ and edge flows $\phi=(\phi_{ij}|(i,j)\in{\cal E})$ corresponding to a given set of gas injections $q=(q_i|i\in{\cal V})$, i.e.:
\begin{eqnarray}
&& \forall (i,j)\in{\cal E}:\quad p_i^2-p_j^2=a_{ij}
{ \phi_{ij}|\phi_{ij}|},
\label{GF1}\\
&& \forall i\in{\cal V}:\quad q_i=\sum_{j:(i,j)\in{\cal E}}\phi_{ij}.\label{GF2}
\end{eqnarray}
We note here that finding a solution to the GF problem in Eqs.~(\ref{GF1}, \ref{GF2}) can be restated as solving a convex optimization \cite{77Nau,12BNV}.

In the steady-state model, the injections are balanced, i.e., $\sum_{i\in{\cal V}} q_i=0$. There is one more node than there are edge equations in (\ref{GF1}), therefore, the pressure must be fixed at one of the nodes. Depending on the structure of the GF network and the gas injections, there may be no physical solution to the GF problem, i.e., the set of feasible solutions to Eqs.~(\ref{GF1}, \ref{GF2}) is an empty set, unless we allow complex values for $p_i$ ($p_i^2 < 0$).
In this case, the GF network cannot support the imposed gas injections and resulting edge flows $\phi_{ij}$ without boosting the pressure with gas compressors.

\begin{figure}
\centering
\includegraphics[width=0.45\textwidth]{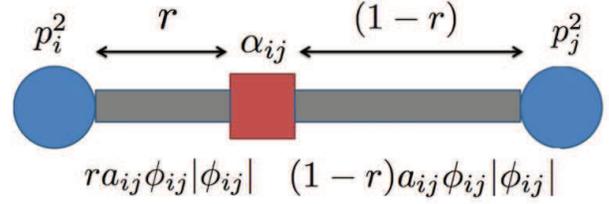}
\caption{Nodes (blue circles), edges (grey line) and compressor (red square) for the gas flow equations in (\ref{GF3b}). The compressor is at relative location $r$ along the edge.  The expressions below the edge are the drops in the square pressures before and after the compressor with compression ratio $\alpha_{ij}$. }
\label{fig:Setup}
\end{figure}

To account for this situation, the GF problem is formulated with compressors placed along edges
$(i,j)$ at a  relative location $r_{ij} \in (0,1)$ (see Fig.~\ref{fig:Setup}).
Let $p_i$ and $p_j$ be
the pressures at nodes $i$ and $j$, respectively.  {
Assuming
positive flow from $i$ to $j$, the compressor inlet square pressure is $p_i^2 - r_{ij}a_{ij}\phi_{ij}|\phi_{ij}|$, and the compressor outlet square pressure is
$p_j^2 + (1-r_{ij})a_{ij}\phi_{ij}|\phi_{ij}|$.  Incorporating flow directions,
equation~(\ref{GF1}) generalizes to, $\forall (i,j)\in{\cal E}:$
\begin{align}
\alpha_{ij} =
\left(\frac{p_j^2 + (1-r_{ij})a_{ij}\phi_{ij}|\phi_{ij}|}
{p_i^2 - r_{ij}a_{ij}\phi_{ij}|\phi_{ij}|}\right)^{\text{sgn}(\phi_{ij})},
\label{GF3b}
\end{align}
where $\alpha_{ij}$ is the ratio of the compressor outlet and inlet square
pressures
along edge $(i,j)$, i.e., the compression ratio (see Fig.~\ref{fig:Setup}).}
$\alpha_{ij}$ is the main control input to the GF network.
For edges without compressors, $\alpha_{ij} = 1$, and
Eq.~(\ref{GF3b}) reduces to Eq.~(\ref{GF1}).  Although a compressor has been added, the flow balance in Eq.~(\ref{GF2}) remains the same.

\subsection{Optimization Problem: Optimum Gas Flow (OGF)}

In the GF model above, the only operational cost is the energy required to run the compressors at compression ratio $\alpha_{ij}$ and mass flux $\phi_{ij}$. We adopt an expression for the cost of compression from \cite{68WL}, i.e.
\begin{eqnarray}
C=\sum_{(i,j)\in{\cal E}} \frac{ c_{ij} |\phi_{ij}|}{\eta_{ij}} { \left( \max\{\alpha_{ij}^m,1\}-1\right)},
\label{cost}
\end{eqnarray}
where $c_{ij}$ is a constant which may depend on the compressor, $\gamma$ is the gas heat capacity ratio, and $0<m=(\gamma-1)/\gamma<1$. $\eta_{ij}$ is the efficiency factor measuring the ratio of the useful power transferred to the gas flow to the shaft power required to run the compressor.  The model in (\ref{cost}) applies to a single compressor or to the aggregate behavior of several identical parallel compressors operating together on a single pipeline
that equally divide the mass flow rate $\phi_{ij}$ over this pipeline\cite{00WRBS}. In (\ref{cost}), we have made a typical assumption that $\eta_{ij}$ is constant. For the configuration of compressors we consider, the most significant deviation from (\ref{cost}) is the dependence of $\eta_{ij}$ on the ratio of the compressor motor speed to the speed of the flow; however, these deviations are relatively small (typically $\sim \pm$5\%) \cite{00WRBS}.  Here, we continue to treat $\eta_{ij}$ as a constant.

Using the cost in (\ref{cost}), the Optimal Gas Flow (OGF) problem is formulated as
\begin{align}
& \min_{\alpha, p,\phi} 
C=\sum_{(i,j)\in{\cal E}} \frac{ c_{ij} {|\phi_{ij}|}}{\eta_{ij}}{ \left( \max\{\alpha_{ij}^m,1\}-1\right)} \label{obj} \\ 
\mbox{s.t.} \quad
& { \forall i \in{\cal V}:\quad q_i = \sum_{j:(i,j)\in{\cal E}} \phi_{ij},}\\
&{  \forall (i,j)\in{\cal E}:\quad \alpha_{ij} =
\left(\frac{p_j^2 + (1-r_{ij})a_{ij}\phi_{ij}|\phi_{ij}|}{p_i^2 -
r_{ij}a_{ij}\phi_{ij}|\phi_{ij}|}\right)^{\text{sgn}(\phi_{ij})},} \\
& \forall i\in{\cal V}:\quad 0\leq \underline{p}_i\leq p_i\leq \overline{p}_i,
\label{p-box}\\
& \forall (i,j)\in{\cal E}: \quad {\underline{\alpha}_{ij}} \leq  \alpha_{ij}\leq \overline{\alpha}_{ij}. \label{alpha-box}
\end{align}
Constraints (\ref{p-box}) come from two different sources.  The upper bound on pressure is an engineering limit defined by the pipeline itself.  The lower bound on pressure is defined by contractual requirements on natural gas delivery pressure.  The upper bound in constraints (\ref{alpha-box}) is another engineering limit on the maximum compression ratio in segment $(ij)$.  The lower bound in constraints (\ref{alpha-box}) is discussed in further detail below.

This OGF formulation differs slightly from previous formulations \cite{68WL}.  In (\ref{alpha-box}), if $\underline{\alpha}_{ij}<1$, we allow both compression and decompression.  Setting  $\underline{\alpha}_{ij} = 1$, eliminates decompression.  Decompression can be implemented by simple procedures (such as a throttling valve) and is usually not associated with any significant cost as it does not require any energy expenditure. In fact, it may be possible to reduce global cost of compression by allowing cost-free decompression at suitable locations. For the sake of completeness, we also address the case when the lower bound in (\ref{alpha-box}) is non-trivial. However, this creates some technical difficulties.  Specifically, the GP that results is non-convex.  In Section~\ref{sec:GP}, we will relax this lower bound and formulate the OGF as a convex geometric program.  In Section~\ref{sec:SP}, we will reintroduce this non-convex lower bound (i.e. $\underline{\alpha}_{ij}=1$) and address it using a Signomial Programming (SP) approach---an approach where the non-convex constraints are linearized and creating an iterative sequence of convex geometric programs.  Results from these two different approaches are discussed in Section~\ref{sec:exp}.

\section{Optimal Gas Flow Algorithms}
\label{sec:tech_opt}

{
\subsection{Tree Network Unique Flow Determination}
\label{sec:tree}

The tree-like topology of pipeline networks guarantees a unique flow solution
in the steady-state.
In general, a steady-state solution can only exist if net injections are
globally balanced, i.e.
$\sum_{i \in {\cal V}} q_i = 0$.
Since the network is a tree, removing any edge $(i,j)$ partitions the network
into two disjoint subgraphs: $\mathcal{G}_i$ and $\mathcal{G}_j$.
Then, the global balance implies that
$\sum_{i \in \mathcal{G}_i} q_i = -\sum_{j \in \mathcal{G}_j} q_j$,
so we must have that
$\phi_{ij} = \sum_{i \in \mathcal{G}_i} q_i = -\sum_{j \in \mathcal{G}_j} q_j$.
In this manner, the flow on every edge of the pipeline tree network can be
uniquely specified.
Thus, for the remainder of this manuscript we treat flow directions and
magnitudes as constants.
This results in the following optimization problem, where $\phi$ is no longer
an optimization variable and $\beta_i = p_i^2$:
\begin{align}
& \min_{\alpha, \beta} \; 
C=
\sum_{(i,j)\in{\cal E}} \frac{ c_{ij} |\phi_{ij}|}{\eta_{ij}}{ \left( \max\{\alpha_{ij}^m,1\}-1\right)} \label{obj2} \\ 
\mbox{s.t.} \quad &
\forall (i,j)\in{\cal E}:\quad \alpha_{ij} =
\left(\frac{\beta_j + (1-r_{ij})a_{ij}\phi_{ij}|\phi_{ij}|}{\beta_i -
r_{ij}a_{ij}\phi_{ij}|\phi_{ij}|}\right)^{\text{sgn}(\phi_{ij})},
\label{alpha_eq}
\\
& 
\forall i\in{\cal V}:\quad 0\leq \underline{\beta}_i\leq \beta_i\leq
\overline{\beta}_i,
\label{p-box2}\\
& 
\forall (i,j)\in{\cal E}:\quad  {\underline{\alpha}_{ij}} \leq  \alpha_{ij}\leq \overline{\alpha}_{ij}. \label{alpha-box2}
\end{align}
}

\subsection{Geometric Programming (GP)}
\label{sec:GP}

Next, we consider the solution of the OGF problem in  (\ref{obj2})-(\ref{alpha-box2}) on a gas network without cycles. The approach is based on Geometric Programming (GP). See \cite{BoydGP}  for a comprehensive discussion of Generalized Geometric Programs (GGPs).

\begin{figure}
\centering
\includegraphics[width=0.45\textwidth]{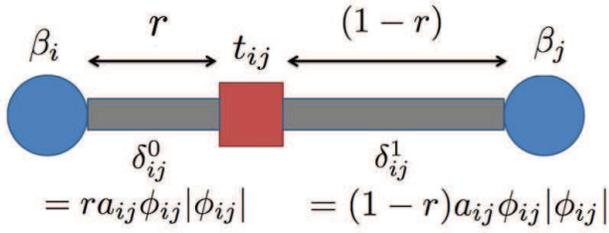}
\caption{Pipeline segment configuration for  the GP formulation.  The color coding of the components is the same as in Fig.~\ref{fig:Setup}. $\delta_{ij}^0$ and $\delta_{ij}^1$ are the drop in the squared pressure $\beta$ from node $i$ to the compressor inlet and from the compressor outlet to node $j$, respectively.  The compression ratio is $t_{ij}$.     }
\label{fig:Tree}
\end{figure}

{
Since flow directions have been determined after solving uniquely for $\phi$,
for simplicity of presentation assume that positive flow is from $i$ to $j$
along every line.  The following derivation applies equally well to the case
when flow is from $j$ to $i$ using Eq.~(\ref{alpha_eq}) for $\alpha$.
Let $d_{ij} = c_{ij}\phi_{ij}/\eta_{ij}$, \;
$\delta_{ij}^0 = r_{ij}a_{ij}\phi_{ij}^2$, and
$\delta_{ij}^1 = (1-r_{ij})a_{ij}\phi_{ij}^2$, which are all constant after solving
for $\phi_{ij}$. Note that $\delta_{ij}^0$ represents the drop in square pressure
from node $i$ to the compressor position while $\delta_{ij}^1$ represents
the drop in square pressure from just after the compressor to node $j$, as
shown in Fig.~\ref{fig:Tree}.
The OGF is then stated as the following optimization problem:
\begin{align}
 	\min_{\alpha,\beta} \quad & \sum_{(i,j) \in \mathcal{E}} d_{ij} (\max \{\alpha_{ij}^{m},1\}-1)  \label{cf} \\
	\mbox{s.t. } \quad &
        \forall i \in \mathcal{V}:\quad 	\underline{\beta}_i \leq \beta_i \leq \bar{\beta}_i, \label{tg_beta} \\
	& \forall (i,j) \in \mathcal{E}:\quad
\alpha_{ij} \leq \bar{\alpha}_{ij},  \label{tg_alpha} \\
& \forall (i,j) \in \mathcal{E}:\quad
\alpha_{ij} = \frac{\beta_j + \delta_{ij}^1}{\beta_i -
\delta_{ij}^0}.
\label{fwd_alpha}
 \end{align}
}
Note that the lower bound constraint on compression  $\underline{\alpha}_{ij}$ is relaxed (\ref{tg_alpha}).  As discussed later, the primary reason for this relaxation is to preserve convexity.  As noted in the previous section, there is a natural justification and operational procedure that corresponds to this relaxation, and we will continue with this relaxation in the rest of Section~\ref{sec:GP}.  To be consistent with the majority of actual operating practices, we will restore this constraint in Section~\ref{sec:SP} and show one way to overcome the technical difficulties it creates.

{
Next we introduce extra variables $t_{ij}$ and rewrite the OGF
\begin{align}
 	\min_{t,\beta} \quad & \sum_{(i,j) \in \mathcal{E}} d_{ij} t_{ij}^m
  \label{cf2} \\
	\mbox{s.t. } \quad &
        \forall i \in \mathcal{V}:\quad 	
\underline{\beta}_i \leq \beta_i \leq \bar{\beta}_i, \label{tg_beta2} \\
	& \forall (i,j) \in \mathcal{E}:\quad
\alpha_{ij} \leq \bar{\alpha}_{ij},  \label{tg_alpha2} \\
& \forall (i,j) \in \mathcal{E}:\quad
\alpha_{ij} = \frac{\beta_j + \delta_{ij}^1}{\beta_i -
\delta_{ij}^0}, \label{alpha_eq2}\\
& \forall (i,j) \in \mathcal{E}:\max\{\alpha_{ij},1\} \leq t_{ij}.
\label{fwd_alpha2}
 \end{align}
Note that the constant term $\sum_{(i,j) \in \mathcal{E}} -d_{ij}$ has been
dropped from the cost function.
Since the cost function is monotonically increasing in $t_{ij}$ for
all $(i,j) \in \mathcal{E}$, at optimum we have
that,
$t_{ij}^{\star} = \alpha_{ij}^{\star}$,
unless $t_{ij}^{\star} = 1$, in which case the current formulation allows
decompression ($\alpha<1$).
Substituting Eq.~(\ref{alpha_eq2}) for $\alpha$ and rearranging
gives
\begin{align}
 	\min_{t,\beta} \quad & \sum_{(i,j) \in \mathcal{E}} d_{ij} t_{ij}^m
  \label{cf2} \\
	\mbox{s.t. } \quad &
        \forall i \in \mathcal{V}:\quad 	
\underline{\beta}_i \leq \beta_i \leq \bar{\beta}_i, \label{tg_beta2} \\
&\forall (i,j) \in \mathcal{E}: 1 \leq t_{ij} \leq \overline{\alpha}_{ij},
\\
& \forall (i,j) \in \mathcal{E}:\frac{\beta_j + \delta_{ij}^1}
{\beta_i - \delta_{ij}^0} \leq t_{ij}.
\label{teq}
\end{align}
The OGF above is equivalent to the following program:
 \begin{align}
 	\min_{t, \beta} \quad & \sum_{(i,j) \in \mathcal{E}} d_{ij} t_{ij}^{m} \label{eq:GP_start}\\
	\mbox{s.t. } \quad & \forall i \in \mathcal{V}:\quad	\underline{\beta}_i \leq \beta_i \leq \bar{\beta}_i,\\
	& \forall (i,j) \in \mathcal{E}:\quad
1 \leq  t_{ij} \leq \bar{\alpha}_{ij}, \\
	& \forall (i,j) \in \mathcal{E}:\quad \beta_j \beta_i^{-1}t_{ij}^{-1} + \delta_{ij}^1 \beta_i^{-1} t_{ij}^{-1} + \delta_{ij}^0 \beta_i^{-1} \leq 1.
 \label{eq:GP_end}
  \end{align}
}

{
This can be reduced to a convex optimization in the form of a geometric program (GP) by introducing variables which are the logarithm of the original variables.
}
Letting $\hat{t}_{ij} = \log t_{ij}$ and $\hat{\beta}_i = \log \beta_i$, we arrive at the convex OGF formulation:
 \begin{align}
 	\min_{\hat{t}, \hat{\beta}} \quad & \log \left( \sum_{(i,j) \in \mathcal{E}} d_{ij} e^{m \hat{t}_{ij}}  \right), \quad \forall i \in \mathcal{V} \label{geombegin} \\
	\mbox{s.t. } \quad &
\forall i \in \mathcal{V}: \quad 	\log(\underline{\beta}_i) \leq \hat{\beta}_i \leq \log(\bar{\beta}_i) \\
	& \forall (i,j) \in \mathcal{E}:\quad
0 \leq  \hat{t}_{ij} \leq \log(\bar{\alpha}_{ij}), \\
	& \forall (i,j) \in \mathcal{E}: \label{geomend}\\
& \log \left( e^{\hat{\beta}_j - \hat{\beta}_i-\hat{t}_{ij}} + \delta_{ij}^1 e^{-\hat{\beta}_i - \hat{t}_{ij}} + \delta_{ij}^0 e^{-\hat{\beta}_i}    \right) \leq 0.   \nonumber
  \end{align}

\subsection{Signomial Programming}
\label{sec:SP}

In current normal practices, pipeline operators do not routinely use decompression as a pressure control.  To be consistent with current operations, the OGF formulation in  Eqs.~(\ref{cf},\ref{tg_beta},\ref{tg_alpha}) is modified by restoring the constraints $1=\underline{\alpha}_{ij}\leq \alpha_{ij}$ for all edges. Note that adding a lower bound of 1 on the compression ratios is the same as $1 \leq \alpha_{ij} =  (\beta_j + \delta_{ij}^1)/(\beta_i - \delta_{ij}^0)$ which after rearranging the terms becomes $\beta_i - \beta_j \leq \delta_{ij}^0 + \delta_{ij}^1$. Following the exact same steps as in the derivation of the GP OGF yields the following optimization:
 \begin{align}
 	\min_{\hat{t}, \hat{\beta}} \quad & \log \left( \sum_{(i,j) \in \mathcal{E}} d_{ij} e^{m \hat{t}_{ij}}  \right) \label{sigbegin} \\
	\mbox{s.t. } \quad & 	\log(\underline{\beta}_i) \leq \hat{\beta}_i \leq \log(\bar{\beta}_i), \quad \forall i \in \mathcal{V} \\
	& 0 \leq  \hat{t}_{ij} \leq \log(\bar{\alpha}_{ij}), \\
	& \log \left( e^{\hat{\beta}_j - \hat{\beta}_i-\hat{t}_{ij}} + \delta_{ij}^1 e^{-\hat{\beta}_i - \hat{t}_{ij}} + \delta_{ij}^0 e^{-\hat{\beta}_i}    \right) \leq 0, \label{sigend1}\\
	& \hat{\beta}_i \leq \log(e^{\hat{\beta_j}} + \delta_{ij} ),  \quad \forall (i,j) \in \mathcal{E} \label{sigend}
  \end{align}
where $\delta_{ij} = \delta_{ij}^0 + {  \delta_{ij}^1}$.

The formulation in (\ref{sigbegin}-\ref{sigend}) is almost a GP, however, the constraints in Eq.~(\ref{sigend}) are non-convex.  We propose to approximately solve (\ref{sigbegin}-\ref{sigend}) with a signomial programming approach---an iterative descent method, where, in each iteration, the non-convex constraints are linearized and the resulting GP is solved to perform one descent step.  The iterations of the algorithm are described below.

\emph{Signomial Programming iteration}
\begin{itemize}
	\item[1.] The constraints Eq.~(\ref{sigend}) are linearized, i.e. $\forall (i,j) \in \mathcal{E}:$
		\begin{eqnarray}
			\hat{\beta_i} \leq \log \left( e^{\hat{\beta}_j^{(t)}} + \delta_{ij} \right) + \frac{e^{\hat{\beta}_j^{(t)}}}{e^{\hat{\beta}_j^{(t)}} + \delta_{ij}} (\hat{\beta}_j  -  \hat{\beta}_j^{(t)}) + \epsilon,\label{linearized}
		\end{eqnarray}
		where a small tolerance parameter $\epsilon > 0$ is added to act as a trade-off between speed of convergence and accuracy.
				
	\item[2.] Solve the Geometric Program that results from Eqs.~(\ref{sigbegin})-(\ref{sigend1}) and Eq.~(\ref{linearized}) to obtain the new iterates at iteration number $t+1$.

    \item[3.] Repeat steps 1 and 2 until the difference in the norms of the solution vectors from one iteration  to the next is less than a specified tolerance $\delta > 0$.
\end{itemize}

The tolerance parameter $\epsilon$ has been introduced to prevent some of the variables from getting frozen at their current value. In particular, for an edge $(i,j) \in \mathcal{E}$ where there is no compressor (i.e.,
$\bar{\alpha}_{ij}  = 1$), we can see that the constraint Eq.~(\ref{sigend1}) reduces to the convex constraint $ \hat{\beta}_i \geq \log(e^{\hat{\beta_j}} + \delta_{ij} )$. In addition, when no decompression is allowed, the above constraint combined with the linearized constraint Eq.~(\ref{linearized}) of the signomial program results in exactly one feasible value for $\hat{\beta_i}$ and $\hat{\beta_j}$. As a result, these variables remain frozen at their initial iterate and this prevents progress in the signomial program. The tolerance parameter $\epsilon$ addresses this issue by allowing a slight violation of the lower bound on the compression ratio, while expanding the feasible region to a neighborhood around the current iterate instead of just one point.

We note that since the constraint Eq.~(\ref{sigend}) is concave, the signomial program outlined above is a special case of the ``concave-convex procedure" \cite{YR03}. It is known that a trust region is
not needed to maintain approximate feasibility in the concave-convex procedure. From the above discussion, we see that the tolerance parameter $\epsilon$ is indeed different from a trust region radius. Smaller tolerance parameters $\epsilon$ and $\delta$ lead to higher accuracy but longer runtimes. If the network consists of a mixture of edges where decompression can be performed and edges where decompression cannot be performed, then the signomial program only needs to linearize the Eq.~(\ref{sigend}) constraints for edges that do not allow decompression. Steps 1 and 2 are repeated until a stopping criterion (3) in the signomial program is reached.

\subsection{Dynamic Programming (DP)} \label{sec:DP}

For comparison of both the formulation and the numerics, we describe a Dynamic Programming (DP) approach to solving the OGF.  The DP approach to OGF is not new.  It was pioneered by \cite{68WL} and has a long history, see e.g. \cite{10Bor} for an extended bibliography. The DP approach exploits the separability of the cost function in Eq.~(\ref{cf}) over the edges as well as the tree structure of the underlying graph by calculating the ``cost-to-go" functions recursively from the leaves upwards.

Specifically, choose a root node (denoted by $r$) for the tree where the pressure is fixed.  At each node $i$, we have a cost-to-go function $J_i(\beta_i)$ which is a function of the squared pressure at that node.  The DP algorithm proceeds as:

(1) \emph{Initialization}.
		Set $\mathcal{S} = \mathcal{V}$, i.e., the set of all nodes.
		For each node $i$ that is a leaf of the tree $\mathcal{G}$ set
		\begin{align*}
			J_i(\beta_i) = \begin{cases}
				0,  \ & \underline{\beta}_i \leq \beta_i \leq \bar{\beta}_i  \\
				\infty, \ & \mbox{otherwise}
			\end{cases}
		\end{align*}
		Remove all the leaves from $\mathcal{S}$.

(2) Repeat the following steps while $\mathcal{S}$ is non-empty:\\
			(a)  Pick a node $i \in \mathcal{S}$ such that all its children have been removed from $\mathcal{S}$.\\
			(b) Let $v_1, \ldots, v_k$ denote the children of $i$. Determine the value of the cost-to-go function $J_i(\beta_i)$ for each $\underline{\beta}_i \leq \beta_i \leq \bar{\beta}_i$ as follows.\\
$\bullet$
					For each choice of compression ratios $\alpha_1, \ldots, \alpha_k$ on the edges $(i,v_1), \ldots, (i, v_k)$ respectively, compute the quantity
						\begin{align*}
							L(\alpha_1, \ldots, \alpha_k) = \sum_{j=1}^{k} d_{i v_j} \alpha_{j}^{m} + J_{v_j}(\beta_{v_j}),
						\end{align*}
						where $\beta_{v_j}$ is the implied squared pressure at $v_j$ for the choice of $\alpha_{j}$ above, i.e.,
						\begin{eqnarray} \label{impliedpressure}
							\beta_{v_j} = \begin{cases} (\beta_i - \delta_{i v_j}^0) \alpha_{j}  - \delta_{i v_j}^1 \quad   &\mbox{if } \phi_{i v_j} > 0, \\
{ (\beta_i + \delta_{v_j i}^1)/\alpha_j + \delta_{v_j i}^0}, \quad &\mbox{otherwise}. \end{cases}
						\end{eqnarray}
$\bullet$						
					Set
						\begin{eqnarray} \label{bellman}
							J_i(\beta_i) = \begin{cases}
								\min\limits_{\alpha_1, \ldots, \alpha_k}  L(\alpha_1, \ldots, \alpha_k)  \  & \mbox{if }  \underline{\beta}_i \leq \beta_i \leq \bar{\beta}_i  \\
								\infty \ & \mbox{otherwise}
							\end{cases}
						\end{eqnarray}
$\bullet$
					Remove $i$ from $\mathcal{S}$.\\		

	(3) \emph{Traceback}. Fix the root squared pressure $\beta_r = \beta_0$ where $\beta_0$ is  the  given squared pressure at the root.
			Set $\mathcal{S} = \mathcal{V}$ to be the set of all nodes. Remove the root $r$ from $\mathcal{S}$.
			Repeat the following while $\mathcal{S}$ is non-empty.\\
				(a) Pick $i \in \mathcal{S}$ such that its parent has been removed from $\mathcal{S}$.\\
                (b) Find the implied pressure $\beta_i$ at $i$ by using the optimal choice of $\alpha$'s in the optimization Eq.~(\ref{bellman}) and using Eq.~(\ref{impliedpressure}).\\
				(c) Remove $i$ from $\mathcal{S}$.
The squared pressures $\beta_i$ obtained in Step 3 are optimal. The optimal value is given by the root cost-to-go function $J_r(\beta_r)$.
In practice for implementation, one needs to discretize the space $\underline{\beta}_i \leq \beta_i \leq \bar{\beta}_i $ for each $i \in \mathcal{V}$ and the space $1 \leq \alpha_{ij} \leq \bar{\alpha}_{ij}$ for each edge
$(i,j) \in \mathcal{E}$ which has a compressor.

\section{Experiments}
\label{sec:exp}

\subsection{Implementation}

The first step for all the algorithms is computing the flow on each edge of the tree networks using explicit expressions for $\phi$ via $q$. Next, we solve and compare the results from several versions of the OGF:  the GP OGF (with relaxed constraints) that allows decompression, the SP OGF that approximates these relaxed constraints, the DP OGF, and a ``greedy compression" scheme that emulates the actions of trained pipeline operators.

Some implementation details:
\subsubsection{Geometric Programming}

The GP OGF is implemented in python using CVXOPT \cite{cvxopt}.

\subsubsection{Signomial Programming}

The GP  iterations of the SP OGF are solved using CVXOPT using the solution of the GP OGF from above as the starting point.

\subsubsection{Dynamic Programming}

The DP OGF was solved using our own code developed in C++ according to the algorithm in Section~\ref{sec:DP}. The number of bins for the $\alpha$'s and $\beta$'s are specified as inputs.  Finer discretization leads to higher accuracy and longer runtime. DP OGF run times increase exponentially with the number of compressors, while signomial programming run times do not.

\subsubsection{Greedy Compression}

A fourth ``greedy compression'' algorithm was implemented for comparison with the GP, SP and DP OGFs. Although exact representation of operator behavior is beyond the scope of this manuscript, we believe this greedy compression algorithm to be a reasonable representation of the day-to-day practice of operators of many natural gas transmission pipelines \cite{Spectra}. Greedy compression is a simple scheme which uses local observations to decide when to compress using the basic rule: {\it whenever the pressure falls below the lower bound, use the nearest upstream compressor to boost the compressor outlet pressure to the maximum value allowed by the local pressure and compression ratio constraints}. However, this simple rule does not always eliminate the violation of pressure constraints.   In this case, a slightly more complicated method is used to select an upstream configuration of compression ratios, however, the decision is still made solely on consideration of local constraint violations.  We omit the details of this selection method for brevity.

\subsection{Models}

We consider two natural gas pipeline networks to test our algorithms--the Belgian gas network \cite{12BNV} and the Transco gas network \cite{Transco} in the Eastern US. Both networks are nearly tree like.  The minor amount of looping in each network was reduced to a tree topology by breaking the loops locations where the flow is expected to be relatively low.  For both test cases, a root node is selected and the square pressure at the root is set to $\overline{\beta}$.

\subsubsection{Belgian Gas Network}

\begin{figure}
\centering
\includegraphics[width=0.49\textwidth]{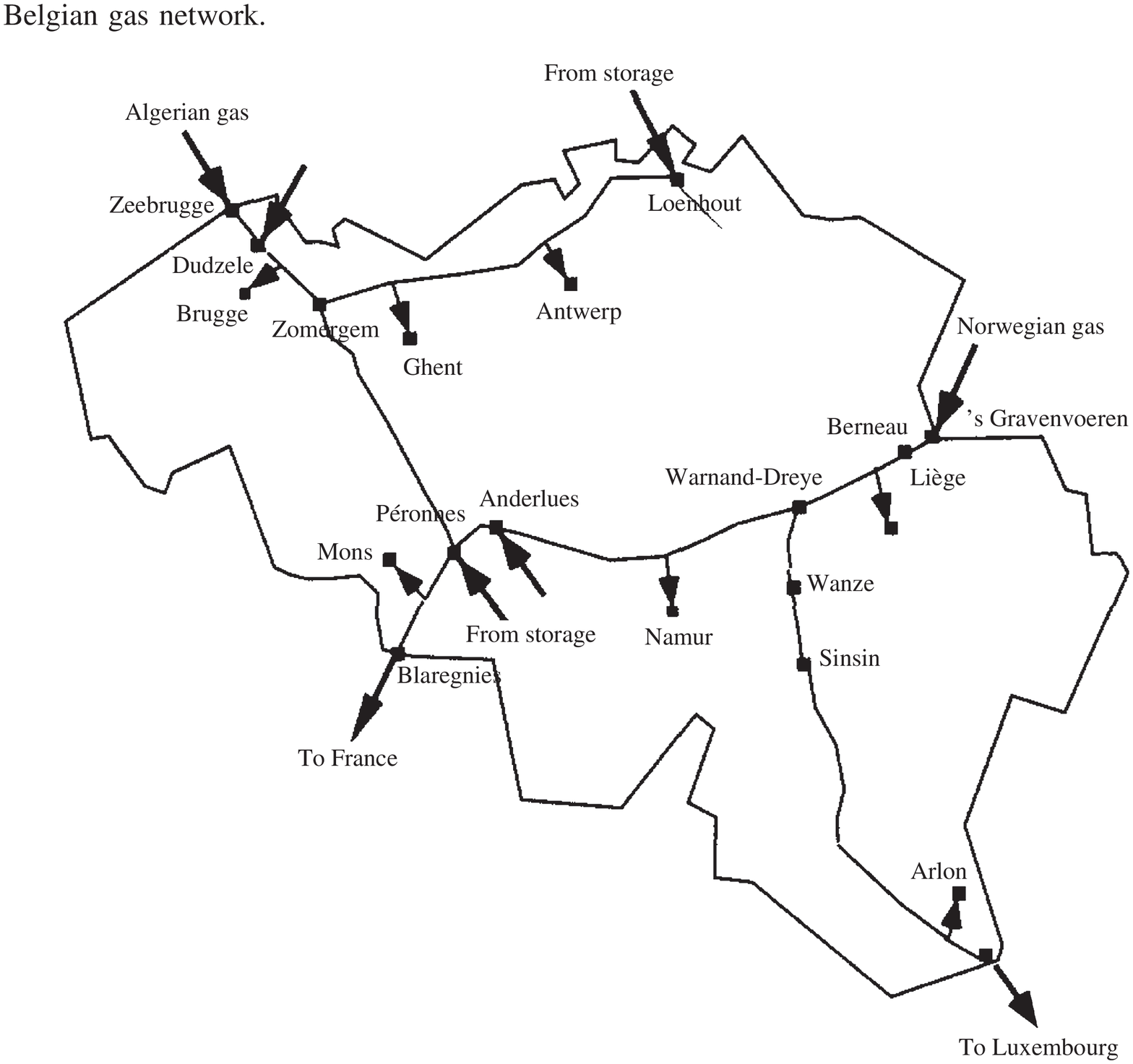}
\caption{Schematic representation of the Belgian gas transmission network.
\label{fig:Belgian}}
\end{figure}

Before comparing the algorithms discussed above on large pipeline networks, we tested the accuracy on a small test case of the Belgian gas network (see Fig.~\ref{fig:Belgian}) and compare our results to those in \cite{12BNV}.  The Belgian network contains 20 nodes and 2 compressors.  Both SP and DP are used to solve for the optimum steady-state compression.  For DP, 1000 pressure bins and 1000 $\alpha$ bins are used.  For SP, $\epsilon$ was set to $10^{-3}$ and the tolerance $\delta$ was set to $10^{-6}$.  Using the same pressure and compression limits as in \cite{12BNV}, the fractional difference between our optimal compression costs and those in \cite{12BNV} is  $\sim 5 \times 10^{-4}$.  Our pressure profiles at optimal compression ratio also agreed with the results in \cite{12BNV}.

To test for the effect of allowing decompression, we compare SP without decompression  and pure GP (which does allow decompression). The fractional difference in optimal costs is $\sim 10^{-2}$ with the geometric programming cost less than for signomial programming.  For this small test case,  the additional freedom of decompression slightly decreases the total cost of compression. In the geometric program solution, decompression of more than $10\%$ was present on $3$ out of $19$ edges in the network.

\subsubsection{Williams Transco Pipeline}

\begin{figure}
\centering
\includegraphics[width=0.45\textwidth]{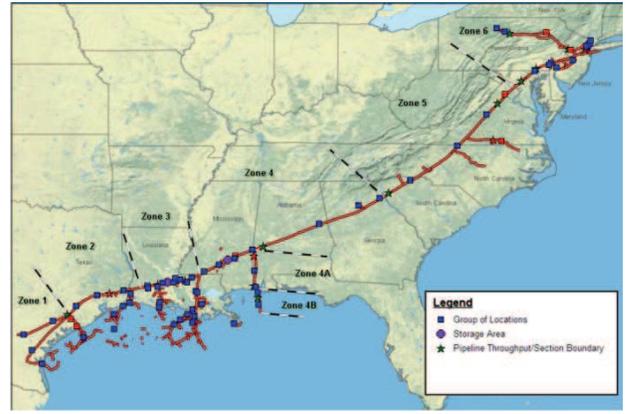}
\caption{Schematic representation of the Transco gas transmission network.  Small loops in the load centers near the northern end of the pipeline were partitioned to create a tree structure.  For this work, the northern half of the pipeline was modeled, starting from the southern border of South Carolina. }
\label{fig:Transco}
\end{figure}

The second and much larger test case is the Willams Transcontinental (Transco) pipeline (see Fig.~\ref{fig:Transco} and \cite{Transco}).  The Transco pipeline extends northeast from gas sources in and around the Gulf of Mexico to load centers in New York and New Jersey.  The structure of the pipeline near to the sources is tree like, however, the details of the gas injections and withdrawls is quite complicated.  Therefore, we choose to test our algorithms on the northern half of the pipeline extending from South Carolina up to the load centers in New Jersey and New York and additional sources in Pennsylvania. We partition a few small loops near the end of the pipeline to achieve a tree-like structure. In spite of reducing the scale of the Transco model, it still consists of 98 nodes and 31
compressors.

The GP, SP and DP algorithms only constrain the pressure at the nodes.  To maintain allowable pressures along the entirety of the pipeline, each compressor segment model has very short runs of inlet and outlet pipeline attached to nodes with zero gas injections.  These short runs of inlet and outlet pipes keep the compressor outlet square pressures from violating $\underline{\beta}$ or $\overline{\beta}$.  The minimum and maximum pressures are set to 500 psi and 800 psi, respectively, as suggested by plots of operational data over this section of pipeline \cite{Transco_report}.

We compare results for signomial programming, dynamic programming, and greedy compression using inflow and injection data from December 29, 2012 \cite{Transo-1line}---near peak load conditions on the Transco pipeline.  For the dynamic programming, 1000 pressure bins and 400 $\alpha$ bins were used.  For the signomial programming, $\epsilon$ is set to
$10^{-2}$ and the tolerance $\delta$ is set to $10^{-3}$.  The fractional difference in optimal costs
between signomial programming and DP is $\sim 3 \times 10^{-5}$.  The greedy compression optimal cost is $5.4\%$ higher than the two other methods demonstrating the benefits of a global optimization approach. The fractional difference between the optimal costs for signomial programming without decompression and pure geometric programming (which does allow decompression) is negligible ($\sim 10^{-7}$,  which is well below tolerance paramters).

Although the GP and SP achieve the same optimal cost, the GP solution involves a significant amount of
decompression.  In particular, $11$ out of $161$ edges show decompression of greater than $10\%$.
Since the cost of decompression is 0, there are often multiple optimal solutions, some of which may contain no decompression. This is the case here.
Inspecting the locations where decompression occurred, we find that most of the decompression occurred at nodes which are along paths that lead to a terminal node when going downstream. 
The pressure bounds at these nodes are well within the upper and lower limits. This happens because the GP solver seems to prefer assigning the minimum pressure at the terminal node, and decompressing (without cost) at edges upstream to achieve this pressure. The SP on the other hand, sets the same compression ratios to one and finds an optimal solution that respects these bounds.

We note here that in this special case where solutions to the GP and SP have the same optimal cost, the SP does not play a significant role. There can be other procedures that can eliminate decompression without changing cost. When the pressure upper and lower bounds are uniform like in our example, some of the optimization variables associated with edges where decompression occurred can in fact be eliminated without consequence. On the other hand, SP will be necessary in networks where there is a difference between the optimal cost between solutions with and without decompression.

\begin{figure}
\centering
\includegraphics[width=0.45\textwidth]{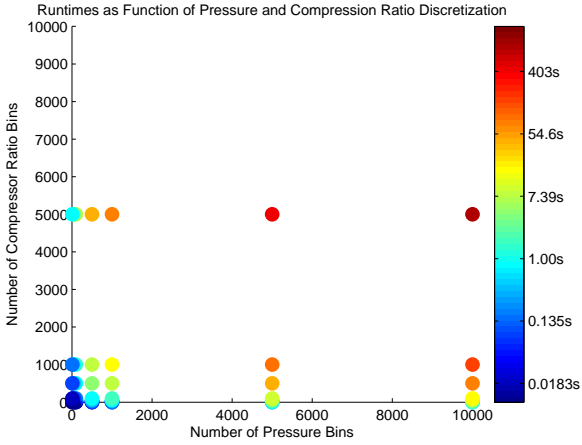}
\caption{Run time versus pressure and compression ratio discretization for the DP solution for the Transco Pipeline. The run time colorbar scales logarithmically in seconds.  {  Run time for the GP solution does not depend on
discretization and is about $5.1$ seconds.}}
\label{fig:Transco_DPruntimes}
\end{figure}

\begin{figure}
\centering
\includegraphics[width=0.45\textwidth]{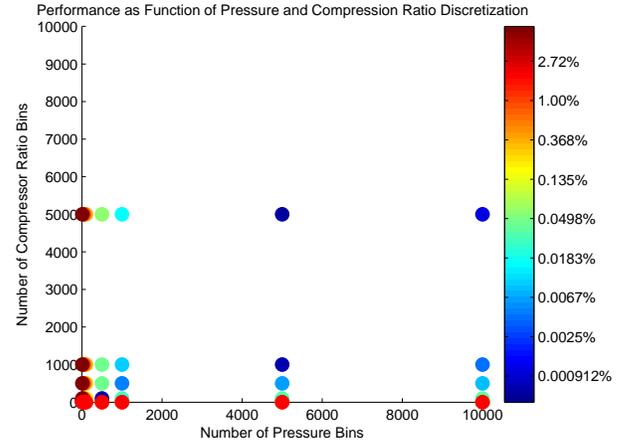}
\caption{Accuracy versus pressure and compression ratio discretization for the DP solution for the Transco Pipeline. The colorbar represents the percent error in the optimal cost.  It scales logarithmically. The bright red dots represent situations where the DP failed. {  The GP solution does not depend on discretization and achieves the correct optimal cost.}}
\label{fig:Transco_DPaccuracy}
\end{figure}

As mentioned earlier, a major advantage of the GP approach is that there is no need for discretization and hence its accuracy is only dependent on the tolerance parameters. On the other hand, bin size and number of bins affect the run time and accuracy of DP significantly. Figures~\ref{fig:Transco_DPruntimes} and \ref{fig:Transco_DPaccuracy} show
plots of the run times and accuracy for the DP OGF for the Transco pipeline, as functions of the number of pressure and compression ratio bins. For a fixed number of pressure bins, the run time scales exponentially with the number of compressor ratio bins.  Similarly, for a fixed number of compressor ratio bins, the run time scales exponentially with the number of pressure bins. For a fixed number of compressor ratio bins, the accuracy tends to scale exponentially with the number of pressure bins. However, for a fixed number of pressure bins, the accuracy does not improve
as the number of compressor bins increases once it has reached some minimum threshold.  {  The GP solution does not depend on discretization and
achieves the correct optimal cost at an average runtime of $5.1$ seconds.}

\begin{figure}
\centering
\includegraphics[width=0.45\textwidth]{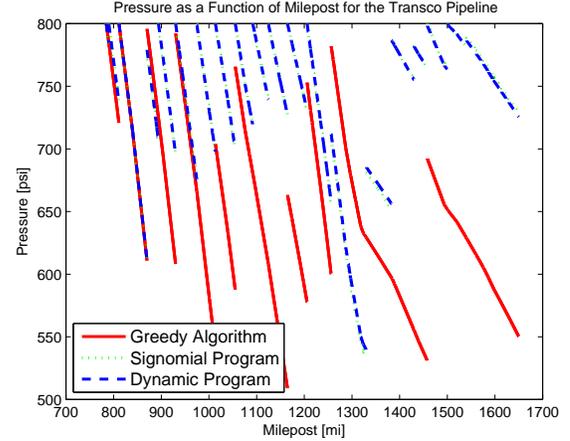}
\caption{Gas pressure versus milepost for SP, DP and greedy compression algorithm solution for the Transco pipeline. }
\label{fig:Transco_pres}
\end{figure}

Fig.~\ref{fig:Transco_pres} shows plots of the pressure as a function
of distance along the pipeline for greedy compression, the SP OGF, and
the DP OGF, respectively.  The SP and DP show negligible differences
while the greedy compression algorithm has a very different pressure
profile. It is interesting to note that, although the greedy algorithm
runs nine compressors in comparison to the nineteen run by the SP OGF
or DP OGF, the cost of compression is higher for the greedy algorithm.
A likely cause for this difference is the lower average gas density,
and therefore higher gas velocities and larger pressure drops, in the
greedy compression case.

\section{Path Forward}
\label{sec:conclusions}

The main contributions of this manuscript are the formulation of the steady-state Optimal Gas Flow (OGF) problem (also called Fuel Cost Minimization Problem in the literature \cite{00WRBS,10Bor}) with a GP approach \cite{BoydGP}---a new approach for this application.  If the lower bound on the compression ratio is relaxed, the OGF on a loop-free gas pipeline network becomes GP that allowing for exact and efficient (polynomial time) solution.  The lower bound on the compression ratio is non-convex, and we show how this constraint can be included using an approximate SP approach.  A significant advantage of the GP and SP methods over the traditional DP approach \cite{68WL,90LP} derives from not having to discretize the node pressure and compression ratio variables.  The GP approach also scales well, even in networks with a high degree of branching whereas the complexity of DP grows exponentially with the degree.

In this manuscript, we made several assumptions based on practical and technical considerations: 1) steady-state gas flow (balanced injections), 2) uniform temperature distribution along the pipe, and 3) the reduction of network cycles to tree-like structures. However, the majority of these assumptions can be relaxed, which form natural extensions to the current work:
\begin{itemize}
\item Many modern gas networks contain no or very few cycles. Combining and extending currently separated (tree-like) systems into one larger and thus more reliable system will lead to the  emergence of significantly meshed systems containing multiple cycles. The extension of the GP approach to the general case of networks with cycles constitutes an interesting challenge. Indeed, finding the flows and finding optimal compression rates --- the two problems which became separable in the tree-network case -- are now mutually dependent. However, this complication can be overcome. One promising approach consists in solving the OGF through multiple repetitions of the following two alternating steps -- (1) finding compression ratios given the flows (where the GP applies directly), and (2) finding flows given compression ratios. Another approach  is to apply the log-change of variables (leading to the convex optimization in the tree case)  followed by relaxation of the new non-convex, cycle-related constraints.

\item Eq.~(\ref{steady}) describes the case of balanced flows, i.e., $\sum_{i\in{\cal V}} q_i=0$. However, this strict balance does not need to hold on the scale of minutes or even hours. When the system is not balanced,  the gas pressure changes leverage the natural storage capacity of pipelines, i.e., linepack.  Exactly accounting for this effect within the basic model described by Eqs.~(\ref{density_eq1},\ref{momenta_eq1}) requires solving a system of coupled PDEs over all pipes of the network \cite{87TT} \cite{94Kiu}, a problem which does not scale well. To achieve a computationally tractable approach,  we plan to approximate Eqs.~(\ref{density_eq1},\ref{momenta_eq1}) with a linearized version. When temporal evolution of sources and sinks is sufficiently slow (so that one can  ignore sound-wave-like transients), the (linearized) diffusive approximation will allow explicit solution for the spatiotemporal and flow dependence of  the pressure, i.e., an approximate solution for the time-dependent line pack and a generalization of Eq.~(\ref{steady}). The result is a generalized OGF that extends what used to be instantaneous optimization into multi-stage optimization that accounts for the evolution of the gas injections over time. We believe the GP approach can be extended to include this temporal evolution.
\end{itemize}

The GP approach has advantages over DP not only because it scales well,  but also because GP allows a fully distributed implementation based on local measurements of pressure and flows at the compressors and local communications between nearest-neighbor compressors. We plan to explore this distributed cyber-physical control \cite{BoydADMM,NedicOzdaglar09} to gas networks in future work.

Finally, this study is motivated by our interest in coupled energy infrastructures, in particular gas and power system networks.  Future increases in stochasticity in one network is expected to have impacts across the other coupled networks.  For example, one mitigation strategy for addressing intermittency of renewable generation, e.g. wind and solar, uses controls on gas turbines to ``smooth'' the  intermittency. However, these gas turbines are loads on the gas network (often burning comparable amount of gas as all other consumers combined).  Therefore, the uncertainty of electric generation translates into temporally fast but spatially long-correlated uncertainty of gas consumption.  Future work will quantify these and other effects of such coupling with a focus on analyzing the stochasticity and correlations across coupled infrastructure networks and using this understanding to develop improved optimization and control of combined systems.

\section*{Acknowledgment}

The authors would like to thank Conrado Borraz-S\'{a}nchez for fruitful discussions and references and Ben Williams of Willams Pipelines for providing customer maps of the Transco pipeline.  The work at LANL was funded by the Advanced Grid Modeling Program in the Office of Electricity in the US Department of Energy and was carried out under the auspices of the National Nuclear Security Administration of the U.S. Department of Energy at Los Alamos National Laboratory under Contract No. DE-AC52-06NA25396.

\bibliographystyle{IEEEtran}
\bibliography{Bib/GasFlow,Bib/Russian,Bib/RefConrado,Bib/GasFlowSM}

\vspace{-1.5cm}
\begin{IEEEbiography}[{\includegraphics[width=1in,height=1.25in,clip,keepaspectratio]{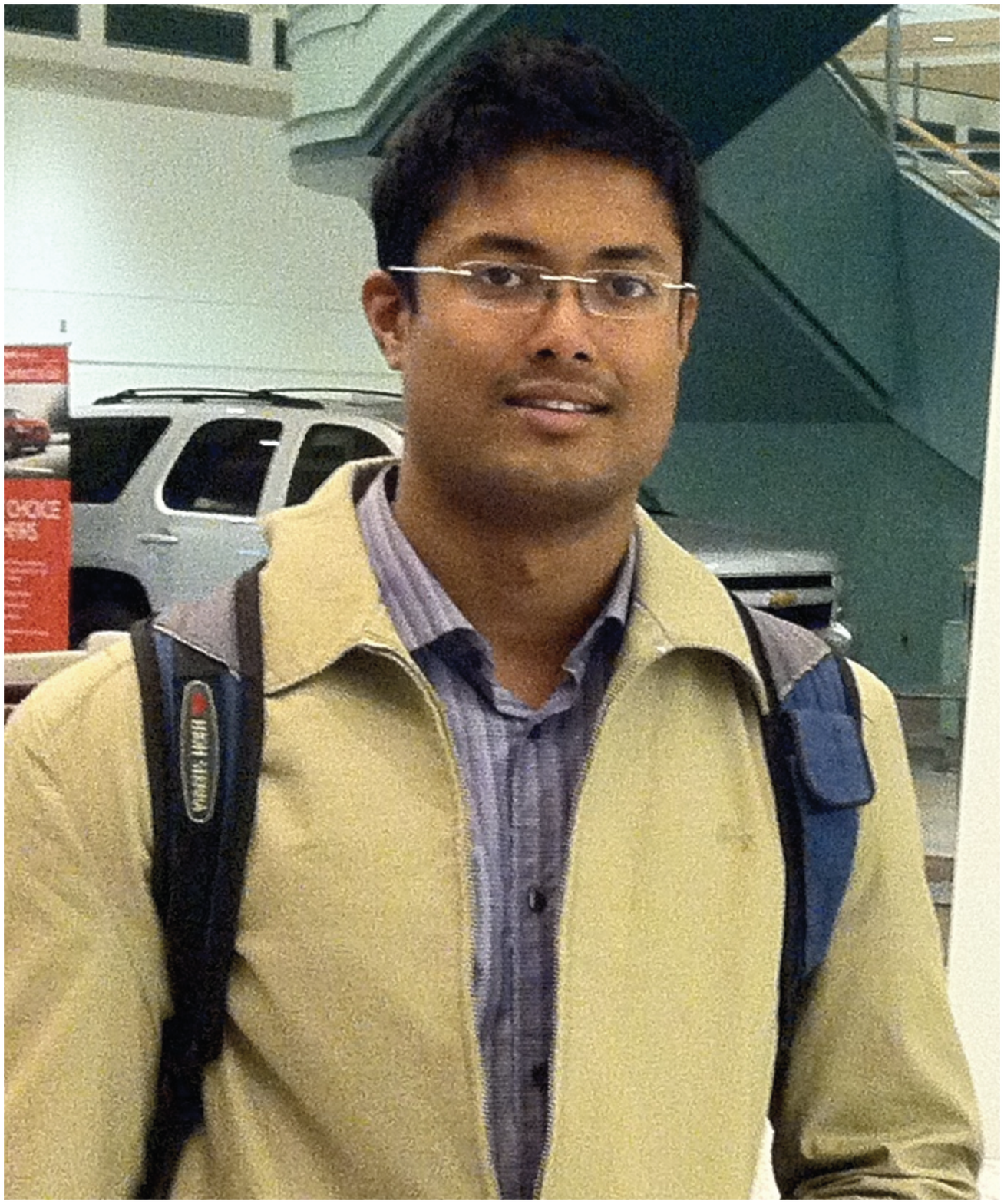}}]{Sidhant Misra}
Sidhant Misra obtained the S.M. and Ph.D. degrees in Electrical Engineering and Computer Science from MIT in 2011 and 2014 respectively.
His research interests include inference and optimization in large scale networks, random graphs and processes and learning in high dimensions, with particular emphasis on message passing algorithms and convex
optimization approaches. He is currently a post-doctoral researcher in the Center for Non-Linear Studies at the Los Alamos National Laboratory.
\end{IEEEbiography}

\vspace{-1.5cm}
\begin{IEEEbiography}[{\includegraphics[width=1in,height=1.25in,clip,keepaspectratio]{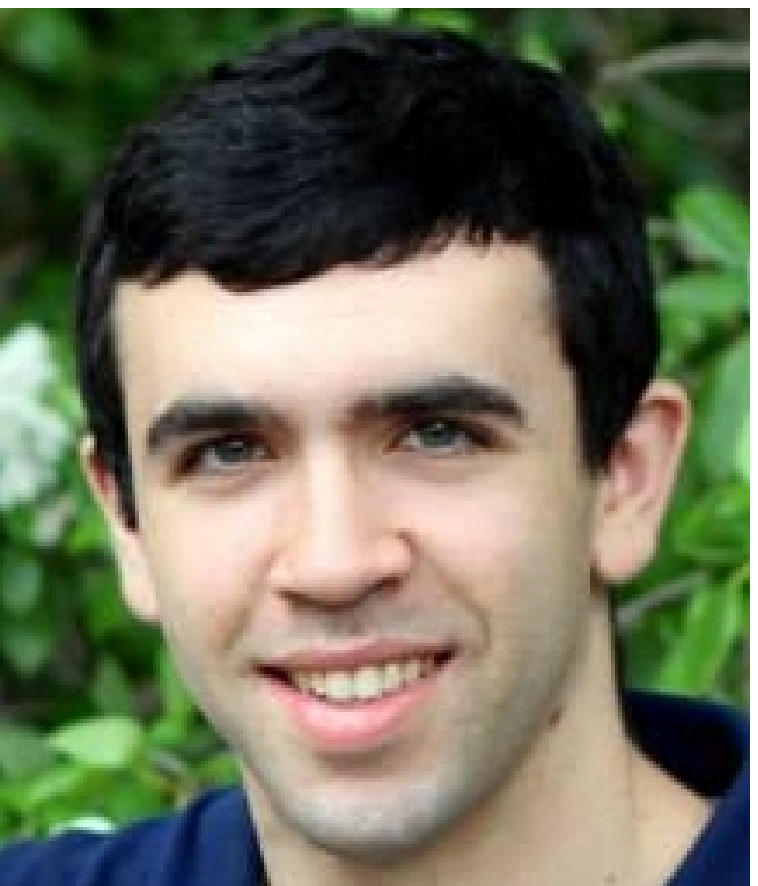}}]{Michael Fisher}
Michael Fisher received his B.A. in physics and mathematics at
Swarthmore College in 2014.  His research interests include
optimization, dynamics, and control of energy systems including power
grids and gas networks.  Currently he is pursuing a Ph.D. in
electrical engineering at the University of Michigan.
\end{IEEEbiography}

\vspace{-1.5cm}
\begin{IEEEbiography}[Photo is not available]
Scott Backhaus received the Ph.D. degree in physics from the University
of California at Berkeley in 1997 in the area of experimental macroscopic
quantum behavior of superfluid He-3 and He-4.
In 1998, he came to Los Alamos, NM, was Director’s Funded
Postdoctoral Researcher from 1998 to 2000, a Reines Postdoctoral
Fellow from 2001 to 2003, and a Technical Staff Member from 2003 to
the present. While at Los Alamos, he has performed both experimental
and theoretical research in the area of thermoacoustic energy conversion
for which he received an R\&D 100 award in 1999 and Technology
Review’s Top 100 Innovators Under 35 [award in 2003]. Recently, his
attention has shifted to other energy-related topics including the fundamental
science of geologic carbon sequestration and grid-integration of
renewable generation.
\end{IEEEbiography}

\vspace{-1.5cm}
\begin{IEEEbiography}[{\includegraphics[width=1in,height=1.25in,clip,keepaspectratio]{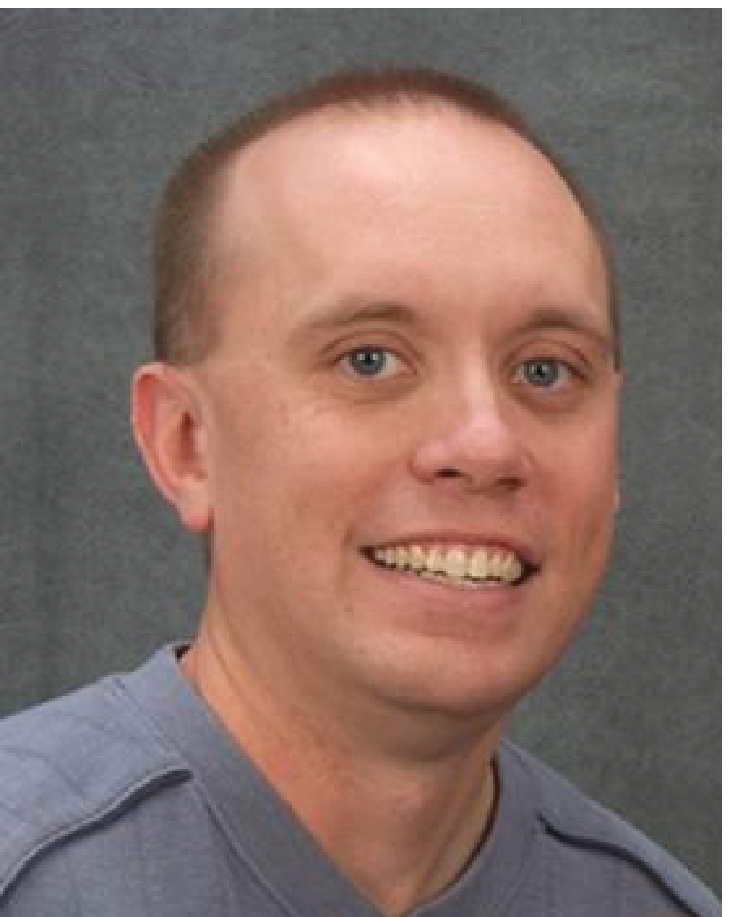}}]{Russell Bent}
Russell Bent is a research scientist in the energy and infrastructure analysis group at Los Alamos National Laboratory. He currently leads research efforts into developing new algorithms for planning, operating, and designing the next generation of critical infrastructure.
His publications include discrete optimization, optimization under uncertainty, infrastructure modeling, constraint programming, and algorithms.
He has published 1 book and over 40 scientific articles.  A full list of
his publications can be found at http://public.lanl.gov/rbent/.
\end{IEEEbiography}

\vspace{-1.5cm}
\begin{IEEEbiography}[{\includegraphics[width=1in,height=1.25in,clip,keepaspectratio]{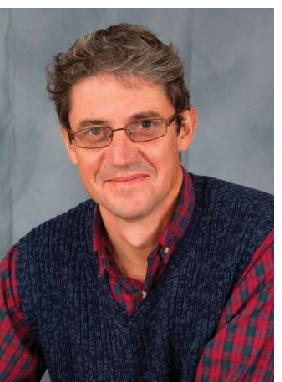}}]{Michael Chertkov}
Dr. Chertkov's areas of interest include statistical and
mathematical physics applied to energy and communication networks,
machine learning, control theory, information theory, computer science,
fluid mechanics and  optics. Dr. Chertkov received his Ph.D. in
physics from the Weizmann Institute of Science in 1996, and his
M.Sc. in physics from Novosibirsk State University in 1990.
After his Ph.D., Dr. Chertkov spent three years at Princeton
University as a R.H. Dicke Fellow in the Department of Physics.
He joined Los Alamos National Lab in 1999, initially as a J.R.
Oppenheimer Fellow in the Theoretical Division. He is now a
technical staff member in the same division. Dr. Chertkov has
published more than 130 papers in these research
areas. He is an editor of the Journal
of Statistical Mechanics (JSTAT), associate editor of IEEE Transactions on
Control of Network Systems, a fellow of the American Physical
Society (APS), and a Founding Faculty Fellow of Skoltech (Moscow, Russia).
\end{IEEEbiography}

\vspace{-15.5cm}
\begin{IEEEbiography}[{\includegraphics[width=1in,height=1.25in,clip,keepaspectratio]{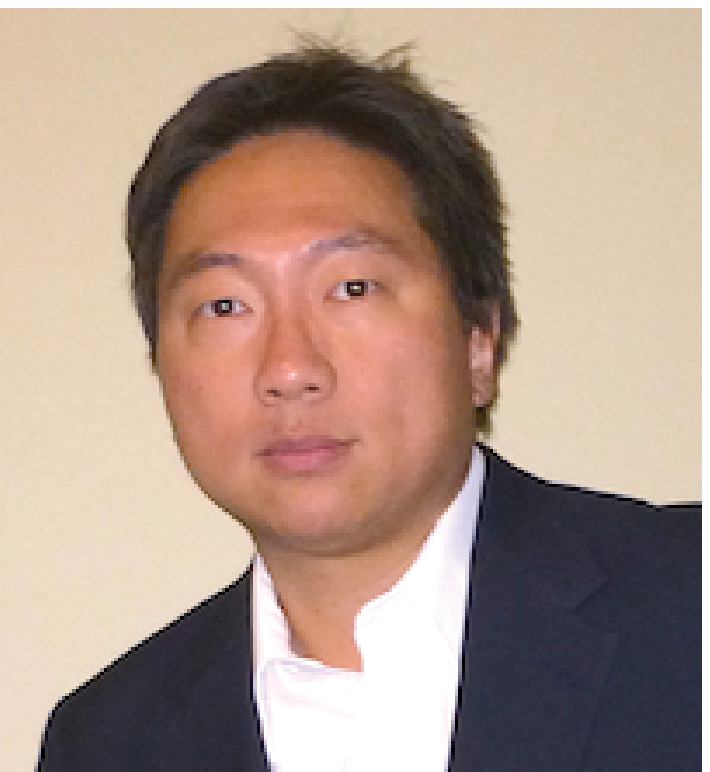}}]{Feng Pan}
Dr. Feng Pan is an engineer in the Electricity Infrastructure Group at the Pacific Northwest National Laboratory.
 He has developed stochastic and network optimization models for energy systems and national security applications. Feng Pan was a  research scientist and project leader in Energy and Infrastructure Analysis Group at the Los Alamos National Laboratory. He received his Ph.D. in Operations Research from the University of Texas at Austin. He served on organizing committees  for INFORMS Annual and society meetings and NSF funded smart grid workshop.
\end{IEEEbiography}

\end{document}